\documentclass[12pt]{iopart}
\usepackage{amsfonts}
\usepackage[english]{babel}
\usepackage{euscript}
\newcommand{\textfrac}[2]{{\textstyle\frac{#1}{#2}}}
\begin{document}

\title[Recursion Operators  for  Integrable rmdKP and rdDym Equations]{%
Recursion Operators and Nonlocal Symmetries
for Integrable rmdKP and rdDym Equations
}

\author{Oleg I. Morozov}

\address{
Institute of Mathematics and Statistics, University of Troms\o, Troms\o 
\, 90-37, Norway
\\ 
E-mail:  Oleg.Morozov{\symbol{64}}uit.no.}

\begin{abstract}
We find direct and inverse recursion operators for integrable cases  of the rmdKP and rdDym equations.
Also, we study actions of these operators on the contact symmetries  and find shadows of nonlocal symmetries of 
these equations.

%\keywords{Lie pseudo-groups \and Maurer--Cartan forms \and symmetries of 
%\and coverings of differential equations}
\end{abstract}


%\vfill \hfill \today

\maketitle

\section{Introduction}

The interrelated notions of infinite hierarchies of symmetries and conservation laws, recursion operators,
bi-Hamiltonian structures, and differential coverings are the main tools in the study of integrable nonlinear
partial differential equations ({\sc pde}s), 
\cite{
Olver1977,
Ovsiannikov1982,
KrasilshchikVinogradov1984,
Ibragimov1985,
KrasilshchikLychaginVinogradov1986,
KrasilshchikVinogradov1989,
Khorkova1989,Kiso1989,
BlumanKumei1989,
Olver1993,
Blaszak1998}.  
In particular, a recursion operator for a 
{\sc pde} is a linear map from the space of symmetries of the {\sc pde} to the same space. Pro\-ce\-du\-res to find 
recursion operators have been proposed by many authors, see, e.g., 
\cite{
Olver1977,
Fuchssteiner1979,
ZakharovKonopelchenko1984,
FokasSantini1986,
Fokas1987,
PapachristouHarrison1988,
FokasSantini1988,
SantiniFokas1988,
BlumanKumei1989, 
Papachristou1991a,
Papachristou1991b,
Olver1993, 
GuthrieHickman1993,
Guthrie1994,
KrasilshchikKersten1994,
KrasilshchikKersten1995,
Marvan1996,
GursesKarasuSokolov1999,
Sergeyev2000,
SandersWang2001,
Wang2002,
MarvanSergyeyev2003,
Marvan2004a,
Marvan2004b,
Sergeyev2005a,
Sergeyev2005b,
ManakovSantini2007,
MarvanPoborzhil2008,
Marvan2010,
PapachristouHarrison2010,
KrasilshchikVerbovetskyVitolo2011,
MalykhSheftel2011,
MarvanSergyeyev2012}. 
As a rule, recursion 
operators are nonlocal. This is one of the reasons motivating the introduction of nonlocal symmetries and, more 
generally, the development of nonlocal geometry of {\sc pde}s, 
\cite{
KrasilshchikVinogradov1984,
KrasilshchikLychaginVinogradov1986,
KrasilshchikVinogradov1989}. In a majority of 
works recursion operators are defined as integro-differential operators, 
\cite{Olver1977,
Fuchssteiner1979,
ZakharovKonopelchenko1984,
FokasSantini1986,
Fokas1987,
FokasSantini1988,
SantiniFokas1988,
BlumanKumei1989, 
Olver1993, 
GursesKarasuSokolov1999,
Sergeyev2000,
SandersWang2001,
Wang2002}, 
although this interpretation  is accompanied by a number of difficulties, e.g., discussed in 
\cite{Guthrie1994,KrasilshchikVinogradov1984}. 
The alternative definition is proposed in \cite{Guthrie1994}, \cite{KrasilshchikKersten1994,KrasilshchikKersten1995}
(see also \cite{PapachristouHarrison2010} and references therein) and developed in 
\cite{
Marvan1996,
Sergeyev2000,
MarvanSergyeyev2003,
Marvan2004a,
Marvan2004b,
MarvanPoborzhil2008,
Marvan2010,
MarvanSergyeyev2012}. This approach considers a recursion operator as an auto-B\"acklund transformation of 
the tangent (or linearized) covering of the {\sc pde}. The machinery of recursion operators become more difficult when 
transitioning from {\sc pde}s with two independent variables to multidimensional {\sc pde}s.  
Accordingly, only a small number of recursion operators for {\sc pde}s in three or more independent variables 
is currently known. In \cite{MarvanSergyeyev2012}, M. Marvan and A. Sergyeyev proposed  the method for 
constructing recursion operators of {\sc pde}s of any dimension from their linear coverings of a special form. 
By this method they found recursion operators for a number of {\sc pde}s of physical and geometrical significance.

In the present paper we consider {\sc pde}s
\begin{equation}
u_{yy} = u_{tx}+u_y u_{xx} - u_x u_{xy}
\label{rmdKP}
\end{equation}
and
\begin{equation}
u_{ty} = u_x u_{xy} - u_y u_{xx}.
\label{rdDym}
\end{equation}
Eq. (\ref{rmdKP}) describes Lorentzian hyperCR Einstein-Weyl structures and is a symmetry re\-duc\-ti\-on of 
Pleba\~nski's second heavenly equation, \cite{Dunajski2004}. It belongs to the family of r-th modified dispersionless 
Kadomtsev--Petviashvili equations (rmdKP) \cite{Blaszak2002},
\[
u_{yy} = u_{tx}+\left(\textfrac{1}{2}\,(\kappa+1)\,u_x^2+u_y\right) u_{xx} +\kappa\, u_x u_{xy}.
\]
Eq. (\ref{rdDym}) is obtained by substituting for $\kappa=-1$ to the family of r-th dispersionless (2+1)-dimensional 
Harry Dym equation (rdDym) \cite{Blaszak2002}, 
\[u_{ty} = u_x u_{xy} +\kappa\, u_y u_{xx}.
\] 
Both Eqs. (\ref{rmdKP}) and (\ref{rdDym}) are known to have two attributes of an integrable {\sc pde}. They are 
bi-Hamiltonian systems on two-dimensional generalizations of the Virasoro algebra, 
\cite{OvsienkoRoger2007,Ovsienko2010}. 
Also, they  have differential coverings  with non-removable parameters. The covering 
\begin{equation}
\left\{
\begin{array}{lcl}
w_t &=& (\lambda^2 - \lambda \, u_x- u_y)\,w_x,
\\
w_y &=& (\lambda - u_x)\,w_x, 
\end{array}
\right.
\label{rmdKP_covering}
\end{equation}
$\lambda \in \mathbb{R}$, for Eq. (\ref{rmdKP}) was found in \cite{Pavlov2003,Dunajski2004}, 
the covering 
\begin{equation}
\left\{
\begin{array}{lcl}
w_t &=& (u_x - \lambda)\, w_x,
\\
w_y &=& \lambda^{-1}\,u_y\, w_x, 
\end{array}
\right.
\label{rdDym_covering}
\end{equation}
of Eq. (\ref{rdDym}) was found for $\lambda = 1$ in \cite{Pavlov2006} and for 
$\lambda \in \mathbb{R} \backslash \{0\}$ in \cite{Morozov2009}.

Also, a hereditary recursion operator for Eq. (\ref{rmdKP}) was found in \cite{ManakovSantini2007}.

In the present paper, we use  the technique of \cite{MarvanSergyeyev2012} to construct recursion operators for Eqs. 
(\ref{rmdKP}) and (\ref{rdDym}). 
Section \ref{Preliminaries_section} is devoted to notation and basic definitions of the geometry {\sc pde}s, 
\cite{Vinogradov1984,
KrasilshchikVinogradov1984,
KrasilshchikLychaginVinogradov1986,
KrasilshchikVinogradov1989,
KrasilshchikVerbovetskyVitolo2011}. 
Section \ref{MS_method} recalls the method of \cite{MarvanSergyeyev2012}. 
In Section  \ref{Recursion_operators_section} we find the direct and inverse recursion operators 
for (\ref{rmdKP}) and (\ref{rdDym}).
In Section \ref{Actions_section} we study actions of these operators on the contact symmetries of (\ref{rmdKP}), 
(\ref{rdDym}) and find shadows of nonlocal symmetries of these equations.

\section{Preliminaries}\label{Preliminaries_section}

Let $\pi \colon \mathbb{R}^n \times \mathbb{R}^m \rightarrow \mathbb{R}^n$,
$\pi \colon (x^1, \dots, x^n, u^1, \dots, u^m) \mapsto (x^1, \dots, x^n)$, be a trivial bundle, and $J^\infty(\pi)$  
be the bundle of its jets of the infinite order. The local coordinates on $J^\infty(\pi)$ are $(x^i,u^\alpha,u^\alpha_I)$, 
where $I=(i_1, \dots, i_n)$ is a multi-index, and for every local section 
$f \colon \mathbb{R}^n \rightarrow \mathbb{R}^n \times \mathbb{R}^m$ of $\pi$ the corresponding infinite jet $j_\infty(f)$ is 
a section $j_\infty(f) \colon \mathbb{R}^n \rightarrow J^\infty(\pi)$ such that
$u^\alpha_I(j_\infty(f)) 
=\displaystyle{\frac{\partial ^{\#I} f^\alpha}{\partial x^I}} 
=\displaystyle{\frac{\partial ^{i_1+\dots+i_n} f^\alpha}{(\partial x^1)^{i_1}\dots (\partial x^n)^{i_n}}}$. 
We put $u^\alpha = u^\alpha_{(0,\dots,0)}$. Also, in the case of $n=3$, $m=1$ we denote $x^1 = t$, $x^2= x$, $x^3= y$, and 
$u^1_{(i,j,k)}=u_{\underbrace{t \dots t}_{i}\underbrace{x \dots x}_{j}\underbrace{y \dots y}_{k}}$.

The  the vector fields
\[
D_{x^k} = \frac{\partial}{\partial x^k} + \sum \limits_{\# I \ge 0} \sum \limits_{\alpha = 1}^m 
u^\alpha_{I+1_{k}}\,\frac{\partial}{\partial u^\alpha_I},
\qquad k \in \{1,\dots,n\},
\]
$(i_1,\dots, i_k,\dots, i_n)+1_k = (i_1,\dots, i_k+1,\dots, i_n)$,  are called {\it total derivatives}.  They commute everywhere on
$J^\infty(\pi)$:  $[D_{x^i}, D_{x^j}] = 0$. 

\vskip 10 pt

The {\it evolutionary differentiation} associated to an arbitrary vector-valued smooth function 
$\varphi \colon J^\infty(\pi) \rightarrow \mathbb{R}^m $ is the vector 
field
\begin{equation}
\mathbf{E}_{\varphi} = \sum \limits_{\# I \ge 0} \sum \limits_{\alpha = 1}^m D_I(\varphi^\alpha)\,\frac{\partial}{\partial u^\alpha_I},
\label{evolution_differentiation}
\end{equation}
with $D_I=D_{(i_1,\dots\,i_n)} =D^{i_1}_{x^1} \circ \dots \circ D^{i_n}_{x^n}$.

\vskip 10 pt

A system of {\sc pde}s $F_r(x^i,u^\alpha_I) = 0$, $\# I \le s$, $r \in \{1,\dots, R\}$ of the order $s \ge 1$ with $R \ge 1$, defines the submanifold 
$\EuScript{E} = \{(x^i,u^\alpha_I) \in J^\infty(\pi) \,\,\vert\,\, D_K(F_r(x^i,u^\alpha_I)) = 0, \,\, \# K \ge 0\}$
in $J^\infty(\pi)$. 

\vskip 10 pt

A function $\varphi \colon J^\infty(\pi) \rightarrow \mathbb{R}^m$ is called a {\it (generator of an infinitesimal) symmetry} of 
$\EuScript{E}$ when $\mathbf{E}_{\varphi}(F) = 0$ on $\EuScript{E}$. The symmetry $\varphi$ is a solution to the {\it defining system} 
\begin{equation}
\ell_{\EuScript{E}}(\varphi) = 0, 
\label{defining_eqns}
\end{equation}
where $\ell_{\EuScript{E}} = \ell_F \vert_{\EuScript{E}}$ with the matrix differential operator 
\[
\ell_F = \left(\sum \limits_{\# I \ge 0}\frac{\partial F_r}{\partial u^\alpha_I}\,D_I\right)
\]
The {\it symmetry algebra} $\mathrm{sym} (\EuScript{E})$ consists of solutions to  (\ref{defining_eqns}). 
The {\it Jacobi bracket} $\{\varphi,\psi\} = \mathbf{E}_{\varphi}(\psi) - \mathbf{E}_{\psi}(\varphi)$ defines a 
structure of a Lie algebra over $\mathbb{R}$ on $\mathrm{sym} \EuScript{E}$.  
The {\it algebra of contact symmetries} $\mathrm{sym}_0 (\EuScript{E})$ is the Lie subalgebra of $\mathrm{sym} (\EuScript{E})$ 
defined as $\mathrm{sym} (\EuScript{E}) \cap J^1(\pi)$.

\vskip 10 pt

A {\it conservation law} of $\EuScript{E}$ is an equivalence class of (n-1)-forms
\[
\omega = \sum \limits_{1\le i_1 < \dots < i_{n-1} \le n} 
b_{i_1\dots i_{n-1}}\,dx^{i_1} \wedge \dots \wedge dx^{i_{n-1}}
\]
with $b_{i_1\dots i_{n-1}} \in C^\infty(J^\infty(\pi))$ such that 
\[
d_h \omega = 
\sum \limits_{k=1}^n \sum \limits_{1\le i_1 < \dots < i_{n-1} \le n} 
D_k(b_{i_1\dots i_{n-1}})\,dx^k \wedge dx^{i_1} \wedge \dots \wedge dx^{i_{n-1}} = 0
\]
on $\EuScript{E}$.  Two such forms are equivalent when their difference is a form
\[
\theta = 
\sum \limits_{k=1}^n \sum \limits_{1\le i_1 < \dots < i_{n-2} \le n} 
D_k(c_{i_1\dots i_{n-2}})\,dx^k \wedge dx^{i_1} \wedge \dots \wedge dx^{i_{n-2}},
\]
so $d_h \theta =0$.

\vskip 10 pt

Denote $\mathcal{W} = \mathbb{R}^\infty$ with  coordinates $w^s$, $s \in  \mathbb{N} \cup \{0\}$. Locally, 
an (infinite-dimensional)  {\it differential covering} of $\EuScript{E}$ is a trivial bundle $\tau \colon J^\infty(\pi) \times \mathcal{W} \rightarrow J^\infty(\pi)$ 
equipped with {\it extended total derivatives} 
\begin{equation}
\widetilde{D}_{x^k} = D_{x^k} + \sum \limits_{ s =0}^\infty
T^s_k(x^i,u^\alpha_I,w^j)\,\frac{\partial }{\partial w^s}
\label{extended_derivatives}
\end{equation}
such that $[\widetilde{D}_{x^i}, \widetilde{D}_{x^j}]=0$ for all $i \not = j$ whenever $(x^i,u^\alpha_I) \in \EuScript{E}$. We define
the partial derivatives of $w^s$ by  $w^s_{x^k} =  \widetilde{D}_{x^k}(w^s)$.  This yields the system of 
{\it covering equations}
\begin{equation}
w^s_{x^k} = T^s_k(x^i,u^\alpha_I,w^j).
\label{WE_prolongation_eqns}
\end{equation}
This over-determined system of {\sc pde}s is compatible whenever $(x^i,u^\alpha_I) \in \EuScript{E}$.

\vskip 10 pt

\noindent
{\sc example} 1.\,\,\,
In the case of $n=3$ and $m=1$ consider the extended total derivatives
\begin{equation}
\left\{
\begin{array}{lcl}
\widetilde{D}_t &=& 
\displaystyle{
D_t + \sum \limits_{k=0}^\infty
\widetilde{D}_x^k ((u_x-\lambda)\,w_1)\,\frac{\partial}{\partial w_k},
}
\\
\widetilde{D}_x &=& 
\displaystyle{
D_x + \sum \limits_{k=0}^\infty
w_{k+1}\,\frac{\partial}{\partial w_k},
}
\\
\widetilde{D}_y &=& 
\displaystyle{
D_y + \lambda^{-1}\,\sum \limits_{k=0}^\infty
\widetilde{D}_x^k (u_y\,w_1)\,\frac{\partial}{\partial w_k}.
}
\end{array}
\right.
\label{rdDym_extended_derivatives}
\end{equation}
Then define the partial derivatives of the fiber variables as $w_{k,t} = \widetilde{D}_t(w_k)$, 
$w_{k,x} = \widetilde{D}_x(w_k)$, and $w_{k,y} = \widetilde{D}_y(w_k)$. This implies
$w_k = w_{0,x\dots x}$ (k times $x$), and
\[
\left\{
\begin{array}{lcl}
w_{k,t} &=& ((u_x-\lambda)\,w_1)_{x\dots x},
\\
w_{k,y} &=& \lambda^{-1}(u_y\,w_1)_{x\dots x}.
\end{array}
\right.
\] 
All these equations are differential  consequences of the system
\[
\left\{
\begin{array}{lcl}
w_{0,t} &=& (u_x-\lambda)\,w_{0,x},
\\
w_{0,y} &=& \lambda^{-1}\,u_y\,w_{0, x}.
\end{array}
\right.
\]
We put $w_0 = w$ and get Eqs. (\ref{rdDym_covering}). Thus the covering with the extended total derivatives
(\ref{rdDym_extended_derivatives}) is defined by (\ref{rdDym_covering}).

\vskip 10 pt

Denote by $\widetilde{\mathbf{E}}_\varphi$  the result of substitution for $\widetilde{D}_{x^k}$ instead of  
$D_{x^k}$ in (\ref{evolution_differentiation}). A  {\it shadow of  nonlocal symmetry} of $\EuScript{E}$ 
corresponding to the covering $\tau$ with the extended total derivatives (\ref{extended_derivatives}), 
or $\tau$-{\it shadow}, is a function $\varphi \in C^\infty(\EuScript{E} \times \mathcal{W})$ such that 
\begin{equation}
\widetilde{\mathbf{E}}_\varphi(F) = 0
\label{shadow_eqn}
\end{equation} 
is a consequence of equations $D_K(F)=0$ and (\ref{WE_prolongation_eqns}).
A {\it nonlocal symmetry} of $\EuScript{E}$ cor\-res\-pon\-ding to the covering $\tau$ (or $\tau$-{\it symmetry}) 
is the vector field
\begin{equation}
\widetilde{\mathbf{E}}_{\varphi,A} = \widetilde{\mathbf{E}}_{\varphi}
+\sum \limits_{s=0}^\infty A^s\,\frac{\partial}{\partial w_s},
\label{extended_nonlocal_evolution_differentiation}
\end{equation}
with $A^s \in C^\infty(\EuScript{E} \times \mathcal{W})$ such that
$\varphi$ satisfies to  (\ref{shadow_eqn})  and
\begin{equation}
\widetilde{D}_k(A^s)=\widetilde{\mathbf{E}}_{\varphi,A}(T^s_k)
\label{obstruction_eqn}
\end{equation}
for $T^s_k$ from (\ref{extended_derivatives}), see \cite[Ch. 6, \S 3.2]{VK1999}.

\vskip 10 pt
\noindent
{\sc remark} 1.  
In general, not every $\tau$-shadow  corresponds to a $\tau$-symmetry, since Eqns. (\ref{obstruction_eqn})
provide an obstruction for existence of (\ref{extended_nonlocal_evolution_differentiation}). But for any $\tau$-shadow
$\varphi$ there exists a covering $\tau_{\varphi}$ and a nonlocal $\tau_{\varphi}$-symmetry whose
$\tau_{\varphi}$-shadow coincides with $\varphi$, see \cite[Ch. 6, \S 5.8]{VK1999}.

\vskip 10 pt

A {\it recursion operator} $\mathcal{R}$   for $\EuScript{E}$ is a $\mathbb{R}$-linear map such that for each  (local or nonlocal)  symmetry $\varphi$ of  $\EuScript{E}$ the function $\mathcal{R}(\varphi)$ is a (local or nonlocal) symmetry of $\varphi$ of  $\EuScript{E}$.

\vskip 10 pt
%=====================================================================================
%=====================================================================================
%=====================================================================================

The tangent covering for {\sc pde} $\EuScript{E}$ is defined as follows, \cite{KrasilshchikVerbovetskyVitolo2011}. 
Consider the trivial bundle
$\sigma \colon J^\infty(\pi)\times \mathcal{Q} \rightarrow J^\infty(\pi)$ 
with  coordinates $q^\alpha_I$, $\# I \ge 0$, on the fibre $\mathcal{Q}$ equipped with the extended total derivatives
\[
\hat{D}_{x^k} = D_{x^k}+\sum \limits_{\# I \ge 0} \sum \limits_{\alpha = 1}^m 
q^\alpha_{I+1_k}\,\frac{\partial}{\partial q^\alpha_I}.
\]
Then for $\hat{D}_I= \hat{D}^{i_1}_{x^1} \circ \dots \circ \hat{D}^{i_n}_{x^n}$ define 
\[
\hat{\ell}_F = \left(
\sum \limits_{\# I \ge 0}\frac{\partial F_r}{\partial u^\alpha_I}\,\hat{D}_I
\right).
\]
and put
\[
\fl
\EuScript{T(E)} = 
\{
(x^i,u^\alpha_i,q^\alpha_I) \in J^\infty(\pi)\times \mathcal{Q}
\,\,\,\vert \,\,\,
D_K(F(x^i,u^\alpha_I))=0,\,\,
\hat{D}_K (\hat{\ell}_F(q^\alpha))=0,\,\,
\# K \ge 0
\}.
\]
The {\it  tangent covering} is the restriction of $\sigma$ to $\EuScript{T(E)}$. A section 
$\varphi \colon \EuScript{E} \rightarrow \EuScript{T(E)}$ of the tangent covering is a symmetry of $\EuScript{E}$.  
The extended total derivatives of this covering are 
$\widetilde{D}_{x^k} = \hat{D}_{x^k}\vert_{\EuScript{T(E)}}$.

\vskip 10 pt 
\noindent
{\sc example} 2.\,\,\,
We write Eq. (\ref{rdDym}) in the form $u_{ty} - u_x\,u_{xy}+u_y\,u_{xx} = 0$. Then we have
\[
\ell_F(\varphi) = D_tD_y(\varphi)-u_xD_xD_y(\varphi)-u_{xy}D_x(\varphi)+u_yD^2_x(\varphi)+u_{xx}D_y(\varphi)
\]
and
\[
\hat{\ell}_F(q) = q_{(1,0,1)}-u_x q_{(0,1,1)}-u_{xy}q_{(0,1,0)}+u_yq_{(0,2,0)}+u_{xx}q_{(0,0,1)}.
\]
The fiber of the tangent covering has local coordinates 
$q_{(i,j,0)}$ and $q_{(0,j,k)}$. The extended total derivatives of the tangent covering are
\[
\fl
\left\{
\begin{array}{lcl}
\widetilde{D}_t &=& 
\displaystyle{
D_t 
+ \sum \limits_{i=0}^\infty \sum \limits_{j=0}^\infty
q_{(i+1,j,0)}\,\frac{\partial}{\partial q_{(i,j,0)}}
+ \sum \limits_{j=0}^\infty 
q_{(1,j,0)}\,\frac{\partial}{\partial q_{(0,j,0)}}
}
\\
&&
\displaystyle{
\phantom{D_t}
+ \sum \limits_{j=0}^\infty \sum \limits_{k=1}^\infty
\tilde{D}_y^{k-1}(u_x q_{(0,1,1)}+u_{xy}q_{(0,1,0)}-u_yq_{(0,2,0)}-u_{xx}q_{(0,0,1)})\,
\frac{\partial}{\partial q_{(0,j,k)}},
}
\\
\widetilde{D}_x &=& 
\displaystyle{
D_x 
+ \sum \limits_{i=0}^\infty \sum \limits_{j=0}^\infty
q_{(i,j+1,0)}\,\frac{\partial}{\partial q_{(i,j,0)}}
+ \sum \limits_{j=0}^\infty \sum \limits_{k=0}^\infty
q_{(0,j+1,k)}\,\frac{\partial}{\partial q_{(0,j,k)}},
}
\\
\widetilde{D}_y &=& 
\displaystyle{
D_y 
+ \sum \limits_{i=1}^\infty \sum \limits_{j=1}^\infty
\tilde{D}_t^{i-1}(u_x q_{(0,1,1)}+u_{xy}q_{(0,1,0)}-u_yq_{(0,2,0)}-u_{xx}q_{(0,0,1)})\,
\frac{\partial}{\partial q_{(i,j,0)}}
}
\\
&&
\displaystyle{
\phantom{D_y}
+ \sum \limits_{j=0}^\infty 
q_{(0,j,1)}\,\frac{\partial}{\partial q_{(0,j,0)}}
+ \sum \limits_{j=0}^\infty \sum \limits_{k=0}^\infty
q_{(0,j,k+1)}\,\frac{\partial}{\partial q_{(0,j,k)}}
}.
\end{array}
\right.
\]

%=====================================================================================
%=====================================================================================
%=====================================================================================

\vskip 10 pt
\noindent
{\sc remark}  2.    Abusing the notation, we write  $D_t^iD_x^jD_y^k(q)$ instead of $q_{(i,j,k)}$ in what follows.

\section{Recursion operators for equations with linear coverings}\label{MS_method}

The technique  of \cite{MarvanSergyeyev2012} is applicable to {\sc pde}s with linear coverings 
defined by covering equations of the form
\begin{equation}
\sum \limits_{i=1}^n A^i_j \, w_{x^i} = 0,
\qquad j \in \{ 1,2 \},
\label{linear_covering}
\end{equation}
with $A^i_j \in C^\infty(J^\infty(\pi))$. The commutativity condition for the corresponding vector fields
\[
X_j = \sum \limits_{i=1}^n A^i_j \, D_{x^i}
\]
coincides with $D_K(F)=0$, $\# K\ge 0$. One of the key elements of the method of \cite{MarvanSergyeyev2012} 
is the vector field
\[
Z = \sum \limits_{i=1}^n \zeta^i \, D_{x^i}
\]
with $\zeta^i \in C^\infty(J^\infty(\pi))$ such that 
\begin{equation}
[X_j, Z] = \sum \limits_{i=1}^n \mathbf{E}_\varphi (A_j^i) \, D_{x^i},
\qquad j \in \{1, 2\},
\label{Z_defining_equations}
\end{equation}
with $\varphi \in C^\infty(J^\infty(\pi))$. The pair of equations (\ref{Z_defining_equations}) give an over-determined 
system for the functions $\zeta^i$. This system is compatible whenever $\varphi$ is a symmetry of $\EuScript{E}$. 
Then we seek for a  set of functions
$a_1$, ... , $a_n$ either from $C^\infty(J^\infty(\pi))$ or from $C^\infty(J^\infty(\pi) \times \mathcal{W})$ such that the function
\begin{equation}
\psi = \sum \limits_{j=1}^n a_j \, \zeta^j
\label{RO_definition}
\end{equation}
is either a local symmetry of $\EuScript{E}$ or a shadow of a nonlocal symmetry of $\EuScript{E}$ corresponding to 
the covering (\ref{linear_covering}). Since $\zeta^j$  depend on $\varphi$, then (\ref{RO_definition}) defines 
a map $\psi  = \mathcal{R}(\varphi)$. This map is a recursion operator for $\EuScript{E}$.

\section{Recursion operators for the the integrable  rmdKP and rdDym equations}
\label{Recursion_operators_section}

\subsection{rmdKP equation}

We apply the  described above method of \cite{MarvanSergyeyev2012} to the covering (\ref{rmdKP_covering}) of Eq. (\ref{rmdKP}).
The straight\-for\-ward com\-pu\-ta\-ti\-on shows that condition (\ref{Z_defining_equations}) holds if, and only if, the 
functions $\zeta_1$ and $\zeta_3$ are solutions to the following over-determined system
\begin{equation}
\fl
\left\{
\begin{array}{lcl}
D_t(\zeta) &=& (\lambda^2-\lambda\,u_x-u_y)\,D_x(\zeta),
\\
D_y(\zeta) &=& (\lambda-u_x)\,D_x(\zeta),
\end{array}
\right.
\label{rmdKP_zeta_1_eqns}
\end{equation}
while $\zeta_2$ satisfies
\begin{equation}
\fl
\left\{
\begin{array}{lcl}
D_t(\zeta_2) &=& (\lambda^2-\lambda\,u_x-u_y)\,D_x(\zeta_2)
+(\lambda\,u_{tx}+u_{ty})\,\zeta_1
+(\lambda\,u_{xx}+u_{xy})\,\zeta_2
\\
&&
+(u_{tx}+u_y\,u_{xx}+(\lambda-u_x)\, u_{xy})\,\zeta_3 -\lambda \, D_x(\varphi) -D_y(\varphi),
\\
D_y(\zeta_2) &=& (\lambda-u_x)\,D_x(\zeta_2) +u_{yx}\,\zeta_1+u_{xx}\,\zeta_2+u_{xy}\,\zeta_3-D_x(\varphi),
\end{array}
\right.
\label{rmdKP_zeta_2_eqns}
\end{equation}
The last system is compatible whenever $\varphi$ is a symmetry of Eq. (\ref{rmdKP}). Then we have

\vskip 10 pt
\noindent
{\sc theorem} 1.   
{\it The function $a_1 \zeta_1+a_2 \zeta_2 + a_3 \zeta_3$ is a nonlocal symmetry of Eq. (\ref{rmdKP}) 
cor\-res\-pon\-ding to the covering (\ref{rmdKP_covering}) whenever it is a $\mathbb{R}$-linear combination of the functions
\[
\psi = (\lambda^2-\lambda\,u_x-u_y)\,\zeta_1+\zeta_2+(\lambda-u_x)\,\zeta_3
\]
and 
\begin{equation}
\eta= \frac{1}{q_x}\,\zeta,
\label{eta_rmdKP}
\end{equation}
where $q$ satisfies (\ref{rmdKP_covering}) and $\zeta$ is a solution to (\ref{rmdKP_zeta_1_eqns}).
}

\vskip 10 pt
\noindent
{\sc remark} 3.   
The coefficient at $\zeta$ in the r.h.s. of (\ref{eta_rmdKP}) satisfies the identity
\[
\widetilde{D}^2_y(\upsilon) = \widetilde{D}_t\widetilde{D}_x(\upsilon)
+u_y\,\widetilde{D}^2_x(\upsilon)
+u_{xx}\,\widetilde{D}_y(\upsilon)
-u_x\,\widetilde{D}_x\widetilde{D}_y(\upsilon)
-u_{xy}\,\widetilde{D}_x(\upsilon).  
\]
Therefore, the function $1/q_x$ is a shadow of a nonlocal symmetry of Eq. (\ref{rmdKP}) corresponding to 
the covering (\ref{rmdKP_covering}).

\vskip 10 pt

Since $\zeta_1$, $\zeta_3$ satisfy (\ref{rmdKP_zeta_1_eqns}) and     
$\zeta_2$ satisfies (\ref{rmdKP_zeta_2_eqns}), then $\psi$ is a solution to the following over-determined system
\begin{equation}
\fl
\left\{
\begin{array}{lcl}
D_t(\psi) &=& (\lambda^2-\lambda\,u_x-u_y)\,D_x(\psi) +(\lambda\,u_{xx}+u_{xy})\,\zeta_2\,\psi
-\lambda \, D_x(\varphi) -D_y(\varphi),
\\
D_y(\psi) &=& (\lambda-u_x)\,D_x(\psi) + u_{xx}\,\psi-D_x(\varphi),
\end{array}
\right.
\label{rmdKP_psi_eqns}
\end{equation}
This system   is compatible whenever $\varphi$ is a symmetry of Eq. (\ref{rmdKP}). Each so\-lu\-ti\-on $\psi$ to this system is a symmetry of (\ref{rmdKP}), too. Therefore, Eqs. (\ref{rmdKP_psi_eqns}) define a recursion operator $\psi=\mathcal{R}(\varphi)$ for (\ref{rmdKP}). 
We express $D_x(\varphi)$ and $D_y(\varphi)$ from (\ref{rmdKP_psi_eqns}):
\begin{equation}
\fl
\left\{
\begin{array}{lcl}
D_x(\varphi) &=& (\lambda-u_x)\,D_x(\psi)-D_y(\psi)+u_{xx}\,\psi,
\\
D_y(\varphi) &=& \lambda\,D_y(\psi)-D_t(\psi)-u_y\,D_x(\psi)+u_{xy}\,\psi.
\end{array}
\right.
\label{rmdKP_phi_eqns}
\end{equation}
This system is compatible whenever $\psi$ is a symmetry of (\ref{rmdKP}). Whence (\ref{rmdKP_phi_eqns}) defines the in\-ver\-se recursion operator $\varphi=\mathcal{R}^{-1}(\psi)$. Both systems (\ref{rmdKP_psi_eqns}) and (\ref{rmdKP_phi_eqns})  define 
B\"acklund trans\-for\-ma\-ti\-ons for the tangent covering of Eq. (\ref{rmdKP}).

\subsection{rdDym equation}

For Eq. (\ref{rdDym}) the computations are very similar to those of the previous subsection. Eqs. (\ref{Z_defining_equations})
yield two systems
\begin{equation}
\fl
\left\{
\begin{array}{lcl}
D_t(\zeta) &=& (u_x-\lambda)\,D_x(\zeta),
\\
D_y(\zeta) &=& \lambda^{-1}\,u_y\,D_x(\zeta),
\end{array}
\right.
\label{rdDym_zeta_1_eqns}
\end{equation}
and
\begin{equation}
\fl
\left\{
\begin{array}{lcl}
D_t(\zeta_2) &=& (u_x-\lambda)\,D_x(\zeta_2)-u_{tx}\,\zeta_1-u_{xx}\,\zeta_2-u_{xy}\,\zeta_3+D_x(\varphi),
\\
D_y(\zeta_2) &=& \lambda^{-1}\,\left( u_y\,D_x(\zeta_2) -(u_x u_{xy} - u_y u_{xx})\,\zeta_1-u_{xy}\,\zeta_2
-u_{yy}\,\zeta_3+D_y(\varphi)\right),
\end{array}
\right.
\label{rdDym_zeta_2_eqns}
\end{equation}
such that $\zeta_1$ and $\zeta_3$ are solutions to (\ref{rdDym_zeta_2_eqns}), while the system for $\zeta_2$ is compatible whenever $\varphi$ is a symmetry of (\ref{rdDym}). Then routine computations give

\vskip 10 pt

\noindent
{\sc theorem} 2.    
{\it 
The function $a_1 \zeta_1+a_2 \zeta_2 + a_3 \zeta_3$ is a nonlocal symmetry of Eq. (\ref{rdDym}) 
cor\-res\-pon\-ding to the covering (\ref{rdDym_covering}) if, and only if, it is a $\mathbb{R}$-linear combination of the functions
\begin{equation}
\psi = (u_x-\lambda)\,\zeta_1+\zeta_2+\lambda^{-1}\,u_y\,\zeta_3
\label{rdDym_psi_definition}
\end{equation}
and 
\begin{equation}
\eta= \frac{1}{q_x}\,\zeta,
\label{eta_rdDym}
\end{equation}
where $q$ and $\zeta$ meet  (\ref{rdDym_covering}) and (\ref{rdDym_zeta_1_eqns}), respectively.

}

\vskip 10 pt
\noindent
{\sc remark} 4.   
The coefficient at $\zeta$ in the r.h.s. of (\ref{eta_rdDym}) is a solution of the following e\-qu\-a\-ti\-on:
\[
\widetilde{D}_t\widetilde{D}_y(\upsilon) = 
u_x\,\widetilde{D}_x\widetilde{D}_y(\upsilon)
+u_{xy}\,\widetilde{D}_x(\upsilon)
-u_y\,\widetilde{D}^2_x(\upsilon)
-u_{xx}\,\widetilde{D}_y(\upsilon).  
\]
Therefore, the function $1/q_x$ is a shadow of a nonlocal symmetry of Eq. (\ref{rdDym}) corresponding to the 
covering (\ref{rdDym_covering}).

\vskip 10 pt

The function (\ref{rdDym_psi_definition})  is a solution to the system
system
\begin{equation}
\left\{
\begin{array}{lcl}
D_t(\psi) &=& (u_x-\lambda)\,D_x(\psi) - u_{xx}\,\psi+D_x(\varphi),
\\
D_y(\psi) &=& \lambda^{-1}\,\left(u_y\,D_x(\psi) - u_{xy}\,\psi+D_y(\varphi)\right),
\end{array}
\right.
\label{rdDym_psi_eqns}
\end{equation}
This auto-B\"acklund transformation for the tangent covering of (\ref{rdDym}) defines the recursion operator
$\psi=\mathcal{R}(\varphi)$. The inverse recursion operator $\varphi=\mathcal{R}^{-1}(\psi)$ is defined by the system
\begin{equation}
\left\{
\begin{array}{lcl}
D_t(\varphi) &=& D_t(\psi) + (\lambda-u_x)\,D_x(\psi) + u_{xx}\,\psi
\\
D_y(\varphi) &=& \lambda\,D_y(\psi) -u_y\,D_x(\psi)+ u_{xy}\,\psi.
\end{array}
\right.
\label{rdDym_phi_eqns}
\end{equation}

\section{Actions of recursion operators on contact symmetries}\label{Actions_section}

We consider actions of the recursion operators (\ref{rmdKP_psi_eqns}), (\ref{rdDym_psi_eqns}) and their inverses 
(\ref{rmdKP_phi_eqns}), (\ref{rdDym_phi_eqns}) to the contact symmetries of the corresponding Eqns. (\ref{rmdKP}) and 
(\ref{rdDym}). The standard com\-pu\-ta\-ti\-o\-nal procedures 
\cite{
Ovsiannikov1982,
Vinogradov1984,
Ibragimov1985,
KrasilshchikLychaginVinogradov1986,
BlumanKumei1989,
Olver1993,
VK1999} 
provide generators $\varphi \in \mathrm{sym}_0(\EuScript{E})$. When $\varphi$ are known, we solve Eqns. (\ref{rmdKP_psi_eqns}) and (\ref{rdDym_psi_eqns})  for $\psi = \mathcal{R}(\varphi)$. 
To find actions of $\mathcal{R}^{-1}$ on contact symmetries, we consider $\psi$ in (\ref{rmdKP_phi_eqns}), (\ref{rdDym_phi_eqns})
to be known elements of $\mathrm{sym}_0(\EuScript{E})$ and then solve these systems for $\varphi$. 
For both Eqns. (\ref{rmdKP}) and (\ref{rdDym}) it appears to be easier to find actions of $\mathcal{R}^{-1}$ than ones of 
$\mathcal{R}$, so we start from $\mathcal{R}^{-1}$ in both cases.

\subsection{rmdkP equation}

The infinitesimal generators of the contact symmetry algebra for Eq. (\ref{rmdKP}) are 
\begin{eqnarray*}
\fl
\varphi_0 (A_0)&=& A_0u_t+\frac{1}{2}\,\left(2\,A_0^{\prime}x+A_0^{\prime\prime}y^2\right)\,u_x+A_0^{\prime}y\,u_y 
-A_0^{\prime}u-A_0^{\prime\prime}x\,y-\frac{1}{6}\,A_0^{\prime\prime\prime}y^3,
\\
\fl
\varphi_1(A_1) &=& A_1^{\prime}y\,u_x+A_1\,u_y - A_1^{\prime}\,x-\frac{1}{2}\,A_1^{\prime\prime}y^2,
\\
\fl
\varphi_2(A_2) &=& A_2 u_x -A_2^{\prime}y,
\\
\fl
\varphi_3(A_3) &=& A_3,
\end{eqnarray*}
\[
\fl
\varphi_4 = 2\,x\,u_x +	y\,u_y-3\,u,
\qquad
\varphi_5 = y\,u_x -2\,x,
\]
where $A_j$ together with  $B_k$ below are arbitrary smooth functions of the variable $t$.
The commutators for these generators read
\[
\{\varphi_j(A_j), \varphi_k(B_k)\} =
\left\{
\begin{array}{lcl}
\varphi_{j+k}(A_j B^{\prime}_k -B_k A^{\prime}_j), & \quad & 0 \le j+k \le 3,
\\
0,&& j+k > 3,
\end{array}
\right.
\]
\[
\{\varphi_k(A_k), \varphi_4\} = k\,\varphi_k(A_k), 
\qquad 0 \le k \le 3,
\]
\[
\{\varphi_0(A_0), \varphi_5\} = \{\varphi_3(A_3), \varphi_5\} = 0,
\]
\[
\{\varphi_1(A_1), \varphi_5\} = \varphi_2(A_1),
\]
\[
\{\varphi_2(A_2), \varphi_5\} = 2\,\varphi_3(A_2),
\]
\[
\{\varphi_4, \varphi_5\} = -\varphi_5.
\]
We denote by 
$\mathfrak{g}_j$ the spans of $\varphi_j(A_j)$ when $A_j \in C^\infty(\mathbb{R})$, $0 \le j \le 3$, and put
$\tilde{\mathfrak{g}}_0 = \mathfrak{g}_0\oplus \mathbb{R}\,\varphi_4$,
$\tilde{\mathfrak{g}}_1 = \mathfrak{g}_1\oplus \mathbb{R}\,\varphi_5$,
$\tilde{\mathfrak{g}}_2 = \mathfrak{g}_2$,
$\tilde{\mathfrak{g}}_3 = \mathfrak{g}_3$.
Then the contact symmetry al\-ge\-bra of (\ref{rmdKP}) $\mathrm{sym}_0(\EuScript{E}) = 
\tilde{\mathfrak{g}}_0 \oplus \tilde{\mathfrak{g}}_1 \oplus \tilde{\mathfrak{g}}_2 \oplus \tilde{\mathfrak{g}}_3$ 
has the following grading
\[
\{\tilde{\mathfrak{g}}_j, \tilde{\mathfrak{g}}_k\} =
\left\{
\begin{array}{lcl}
\mathfrak{g}_0, && j+k = 0,
\\
\tilde{\mathfrak{g}}_{j+k}, & \quad & 0 < j+k \le 3,    
\\
0,&& j+k >3.
\end{array}
\right.
\]

The solutions to Eqns. (\ref{rmdKP_phi_eqns}) are defined up to adding their arbitrary solution with 
$\psi=0$, i.e., an arbitrary element of the subalgebra $\mathfrak{g}_3$.  We will not write these elements explicitly, or, 
in other words, we will consider factor spaces w.r.t. $\mathfrak{g}_3$. For 
$\psi = \varphi_j(A_j)$ with $1 \le j \le 3$ and $\psi=\varphi_5$ we get local solutions
\[
\mathcal{R}^{-1}(\varphi_1(A_1)) = -\varphi_0(A_1)+\lambda\,\varphi_1(A_1),
\]
\[
\mathcal{R}^{-1}(\varphi_2(A_2)) = -\varphi_1(A_2)+\lambda\,\varphi_2(A_2),
\]
\[
\mathcal{R}^{-1}(\varphi_3(A_3)) = \varphi_2(A_3),
\]
\[
\mathcal{R}^{-1}(\varphi_5) = -\varphi_4+\lambda\,\varphi_5.
\]
Further, we have
\begin{eqnarray}
\fl
\mathcal{R}^{-1}(\varphi_0(A_0)) &=& s_0 
+\frac{1}{24}y^4\,A_0^{(iv)}+\frac{1}{6}\,y^2\,(3\,x-y(u_x+\lambda))\,A_0^{\prime\prime\prime}
\nonumber
\\
\fl
&&
+\frac{1}{2}\,\left(y\,(\lambda\,y-2\,x)\,u_x-y^2\,u_y+x^2-2\,\lambda\,x\,y)\right)\,A_0^{\prime\prime}
\nonumber
\\
\fl
&&
-(y\,u_t+(u-\lambda\,x)\,u_x+(x-\lambda\,y)\,u_y+2\,u)\,A_0^{\prime}+\lambda\,u_t\,A_0,
\label{phi_0_rmdKP}
\end{eqnarray}
where $s_0$ is a solution to the following compatible system 
\begin{equation}
\fl
\left\{
\begin{array}{lcl}
s_{0,x} &=& y\,u_x A_0^{\prime\prime}
+(u_x^2+u_y)\,A_0^{\prime}
-(u_{ty}+u_xu_{tx}-u_tu_{xx})\,A_0,
\\
s_{0,y} &=& 
(y\,u_y+u)\,A_0^{\prime\prime}
+(u_t+u_x u_y)\,A_0^{\prime}
-(u_{ty}+u_tu_{xy}-u_yu_{xx})\,A_0.
\end{array}
\right.
\label{rmdKP_s_0_equation}
\end{equation}
In other words, $s_0$ is a potential of the conservation law
\begin{eqnarray*}
\fl
ds_0 \wedge dt&=& 
((y\,u_y+u)\,A_0^{\prime\prime}
+(u_t+u_x u_y)\,A_0^{\prime}
-(u_{ty}+u_tu_{xy}-u_yu_{xx})\,A_0)\,dx \wedge dt
\\
\fl
&&
+
(y\,u_x A_0^{\prime\prime}
+(u_x^2+u_y)\,A_0^{\prime}
-(u_{ty}+u_xu_{tx}-u_tu_{xx})\,A_0)\,dy \wedge d t
\end{eqnarray*}
of Eq. (\ref{rmdKP}). For $\varphi_4$ we get
\begin{equation}
\mathcal{R}^{-1}(\varphi_4) = 4\,s_4  -y\,u_t+(2\,\lambda\,x - 3\,u)\,u_x+(\lambda\,y-2\,x)\,u_y-3\,\lambda\,u,
\label{phi_4_rmdKP}
\end{equation}
where $s_4$ is a solution to 
\begin{equation}
\left\{
\begin{array}{lcl}
s_{4,x} &=& u_y+u_x^2,
\\
s_{4,y} &=& u_t+u_xu_y,
\end{array}
\right.
\label{rmdKP_s_4_equation}
\end{equation}
that is,  a potential of the conservation law
\[
ds_4 \wedge dt = 
(u_t+u_xu_y)\, dx \wedge dt
+(u_y+u_x^2)\, dy \wedge dt.
\]
Thus the action of $\mathcal{R}^{-1}$ to $\varphi_0(A_0)$ and $\varphi_4$ provides shadows of nonlocal symmetries 
to Eq. (\ref{rmdKP}): the infinite set (\ref{phi_0_rmdKP}) corresponds to the covering (\ref{rmdKP_s_0_equation}), 
and  (\ref{phi_4_rmdKP}) corresponds to the covering (\ref{rmdKP_s_4_equation}).

\vskip 10 pt

Factorizing w.r.t. solutions of Eqs. (\ref{rmdKP_psi_eqns}) with $\varphi=0$, we have
\[
\mathcal{R} (\varphi_1(A_1))  = -\varphi_2(A_1) +\lambda\,\varphi_3(A_1),
\]
\[
\mathcal{R} (\varphi_2(A_2))  = \varphi_3(A_2),
\]
\[
\mathcal{R} (\varphi_3(A_3))  = 0,
\]
while the solutions $\mathcal{R} (\varphi_0(A_0))$, $\mathcal{R} (\varphi_4)$, $\mathcal{R} (\varphi_5)$ 
of (\ref{rmdKP_psi_eqns}) with $\varphi=\varphi_0(A_0)$, $\varphi=\varphi_4$, and $\varphi=\varphi_5$, 
respectively, are shadows of new nonlocal symmetries of Eq. (\ref{rmdKP}).

\subsection{rdDym equation}

The Lie algebra of contact symmetries of Eq. (\ref{rdDym}) is generated by
\[
\varphi_0(A_0) = 
u_t\, A_0+x\,u_x A_0^{\prime}-u\,A_0^{\prime}+\frac{1}{2}\,x^2\,A_0^{\prime\prime},
\]
\[
\varphi_1(A_1) = 
u_x\,A_1+x\,A_1^{\prime},
\]
\[
\varphi_2(A_2) = 
A_2,
\]
\[
\varphi_3(B_1) = 
u_y\,B_1,
\]
\[
\varphi_4 = x\,u_x-2\,u,
\]
where  $A_j$ and $C_k$ below are arbitrary functions of $t$, and $B_j$ are arbitrary functions of $y$.
The generators have the following commutators:
\[
\{\varphi_j(A_j), \varphi_k	(C_k)\} =
\left\{
\begin{array}{lcl}
\varphi_{j+k}(A_j C^{\prime}_k -C_k A^{\prime}_j), & \quad & 0 \le j+k \le 2,
\\
0,&& j+k >2,
\end{array}
\right.
\]
\[
\{\varphi_j(A_j), \varphi_3(B_1)\} = 0, 
\qquad 0 \le j \le 2,
\]
\[
\{\varphi_k(A_k), \varphi_4\} = k\,\varphi_k(A_k), 
\qquad 0 \le k \le 2,
\]
\[
\{\varphi_3(B_1), \varphi_3(B_2)\} = \varphi_3(B_1B_2^{\prime}-B_2B_1^{\prime}), 
\]
\[
\{\varphi_3(B_1), \varphi_4\} = 0 
\]
With
$\mathfrak{g}_j = \{\varphi_j(A_j) \,\,\,\vert\,\,\, A_j \in C^\infty(\mathbb{R})\}$,
$0 \le j \le 2$,
$\mathfrak{h} = \{\varphi_3(B) \,\,\,\vert\,\,\, B \in C^\infty(\mathbb{R})\}$,
$\tilde{\mathfrak{g}}_0 = \mathfrak{g}_0\oplus \mathbb{R}\,\varphi_4$,
$\tilde{\mathfrak{g}}_1 = \mathfrak{g}_1$,
$\tilde{\mathfrak{g}}_2 = \mathfrak{g}_2$,
and 
$\mathfrak{g} = \tilde{\mathfrak{g}}_0 \oplus \tilde{\mathfrak{g}}_1 \oplus \tilde{\mathfrak{g}}_2$
we have
$\mathrm{sym}_0(\EuScript{E}) = \mathfrak{g} \oplus \mathfrak{h}$,
$\{\mathfrak{g}, \mathfrak{h}\} =0$,
$\{\mathfrak{h}, \mathfrak{h}\} =\mathfrak{h}$,
while the subalgebra $\mathfrak{g}$ has the following grading
\[
\{\tilde{\mathfrak{g}}_j, \tilde{\mathfrak{g}}_k\} =
\left\{
\begin{array}{lcl}
\mathfrak{g}_0, && j+k = 0
\\
\tilde{\mathfrak{g}}_{j+k}, & \quad & 0 < j+k \le 2,
\\
0,&& j+k >2.
\end{array}
\right.
\]

Then up to adding arbitrary solutions of (\ref{rdDym_psi_eqns}) we have
\[
\mathcal{R}^{-1}(\varphi_1(A_1)) = \varphi_0(A_1)+\lambda\,\varphi_1(A_1),
\]
\[
\mathcal{R}^{-1}(\varphi_2(A_2)) = \varphi_1(A_2),
\]
\[
\mathcal{R}^{-1}(\varphi_3(B_1)) = \lambda\,\varphi_3(B_1),
\]
while 
\begin{eqnarray}
\fl
\mathcal{R}^{-1}(\varphi_0(A_0)) &=& s_0 +\frac{1}{6}\,x^3\,A_0^{\prime\prime\prime}
+\frac{1}{2}\,\,(x\,u_x-2\,u+\lambda\,x)\,A_0^{\prime\prime}
\nonumber
\\
\fl
&&
+(x\,(u_t+\lambda\,u_x)-u\,(u_x+\lambda))A_0^{\prime}
+u_t\,(u_x+\lambda)\,A_0+A_3,
\label{phi_0_rdDym}
\end{eqnarray}
where $s_0$ is a solution of the system
\begin{equation}
\left\{
\begin{array}{lcl}
s_{0,x} &=& A_0\,(u_{tt}-2\,u_x u_{tx})-A_0^{\prime}\,(u_t-u_x^2) ,
\\
s_{0,y} &=& A_0^{\prime}\,u_x u_y - A_1\,(u_y u_{tx}+u_x^2u_{xy}-u_xu_yu_{xx}),
\end{array}
\right.
\label{rdDym_s_0_equation}
\end{equation}
that is a potential of the conservation law
\begin{eqnarray*}
%\fl
ds_0 \wedge dt &=& (A_0^{\prime}\,u_x u_y - A_1\,(u_y u_{tx}+u_x^2u_{xy}-u_xu_yu_{xx})) dy \wedge dt
\\
%\fl
&&
+ (A_0\,(u_{tt}-2\,u_x u_{tx})-A_0^{\prime}\,(u_t-u_x^2))\,dx \wedge dt,
\end{eqnarray*}
and
\begin{equation}
\mathcal{R}^{-1}(\varphi_4) = 3\,s_4+x\,u_t-2\,u\,u_x-2\,\lambda\,u
\label{phi_4_rdDym}
\end{equation}
where $s_4$ meets 
\begin{equation}
\left\{
\begin{array}{lcl}
s_{4,x} &=& u_x^2-u_t,
\\
s_{4,y} &=& u_x u_y,
\end{array}
\right.
\label{rdDym_s_4_equation}
\end{equation}
and therefore defines a conservation law 
\[
ds_4 \wedge dt = (u_x^2-u_t)\, dx \wedge dt + u_x u_y\, dy \wedge dt
\]
of (\ref{rdDym}).  Whence Eq. (\ref{rdDym}) has the infinite set of shadows of nonlocal symmetries  
(\ref{phi_0_rdDym}) corresponding to the covering (\ref{rdDym_s_0_equation}) 
and (\ref{phi_4_rdDym}) corresponding to the covering (\ref{rdDym_s_4_equation}).
Also, factorizing w.r.t. arbitrary solutions of (\ref{rdDym_phi_eqns}) we have
\[
\mathcal{R}(\varphi_0(A_0)) = \varphi_1(A_0)-\lambda\,\varphi_2(A_0),
\]
\[
\mathcal{R}(\varphi_1(A_1)) = \varphi_2(A_1),
\]
\[
\mathcal{R}(\varphi_2(A_2)) = 0,
\]
\[
\mathcal{R}(\varphi_3(B_1)) = \lambda^{-1}\,\varphi_3(B_1),
\]
while the solution $\mathcal{R}(\varphi_4)$ to (\ref{rdDym_phi_eqns}) with $\varphi=\varphi_4$ 
is a shadow of  new nonlocal symmetry of Eq. (\ref{rdDym}).

\section{Conclusion}
In this paper we used the construction of \cite{MarvanSergyeyev2012} to find recursion operators for integrable 
cases of the rmdKP and rdDym equations. As a byproduct of computations, we found shadows of nonlocal symmetries of 
these equations corresponding to their coverings with non\-re\-mo\-va\-ble parameters (Remarks 3 and 4). Also, we 
studied actions of the recursion ope\-ra\-tors to contact symmetry algebras  of (\ref{rmdKP}), (\ref{rdDym}) and 
found shadows of nonlocal symmetries cor\-re\-s\-pon\-d\-ing to coverings generated by conservation laws.  
As it is noted in Remark 1, every shadow $\varphi$ provides a nonlocal symmetry in the corresponding covering 
$\tau_\varphi$ of (\ref{rmdKP}) and (\ref{rdDym}). The structure of the spaces of nonlocal symmetries of Eqns. 
(\ref{rmdKP}) and (\ref{rdDym}) is a subject of a further study. Another pro\-mi\-sing field of research is an 
application of the useful method of \cite{MarvanSergyeyev2012} to other {\sc pde}s with linear coverings which 
have non-removable parameters. Also, the problem of finding recursion operators for nonlinear coverings of 
{\sc pde}s in more than two independent va\-ri\-ab\-les seems to be very interesting and important.

\section*{Acknowledgments}
 I am very grateful to Professors I.S. Krasil${}^{\prime}$shchik, M. Marvan, A.G. Sergyeyev, and M.V. Pavlov for 
 en\-ligh\-t\-e\-n\-ing and  fruitful  discussions.

\end{document}